\documentclass[prd, twocolumn, nofootinbib]{revtex4}
\usepackage{hyperref}
\usepackage{amsfonts,amssymb,
}
\usepackage{exscale,bbm}  
\usepackage{psfrag}
\usepackage[dvips]{graphicx}
\usepackage[all,knot]{xy}  
\usepackage{color}
\usepackage{pdfcolmk}
\usepackage{subfig}
\usepackage{bm}

\newcommand\nn{\nonumber}
\newcommand\beq{\begin{equation}}
\newcommand\eeq{\end{equation}}
\newcommand\be{\begin{eqnarray}}
\newcommand\ee{\end{eqnarray}}
\newcommand\beqa{\begin{eqnarray}}
\newcommand\eeqa{\end{eqnarray}}
\newcommand\ba{\begin{eqnarray}}
\newcommand\ea{\end{eqnarray}}
\newcommand\bean{\begin{eqnarray*}}
\newcommand\eean{\end{eqnarray*}}

\newcommand{\cA}{{\mathcal A}}

\newcommand{\cF}{{\mathcal F}}

\newcommand{\cH}{{\mathcal H}}

\newcommand{\cP}{{\mathcal P}}

\newcommand{\cS}{{\mathcal S}}

\newcommand\U{{\mathrm U}}
\newcommand\C{{\mathbb C}}
\newcommand\N{{\mathbb N}}
\newcommand\R{{\mathbb R}}
\newcommand\Z{{\mathbb Z}}

\renewcommand{\H}{{\mathcal{H}}}

\newcommand{\e}{{\mathsf{e}}}

\def\inv{{\mbox{\tiny -1}}}
\def\minus{{\mbox{\small -}}}

\newcommand\unit{\mathbbm{1}}

\newcommand{\SO}{{\rm SO}}

\newcommand{\su}{\mathfrak{su}}
\newcommand{\SU}{{\rm SU}}
\newcommand{\SL}{{\rm SL}}

\newcommand\acts\,

\def\md{\mbox{d}}

\def\vphi{\varphi}
\def\vphihat{\widehat{\varphi}}

\def\extd{\mathrm {d}}

\newcommand\maps{\colon}

\newenvironment{proof.within.proof}
{\noindent{\it Proof:}}{
\hfill $\Box$ \medskip}

\newcommand\add{{\rm \bf add}  }
\newcommand\sub{{\rm \bf sub}  }
\newcommand\inve{{\rm \bf inv}  }
\newcommand\Tr{{\rm Tr}}

\newcommand{\q}{\quad}

\begin{document}

\title{Non-commutative flux representation for loop quantum gravity}
\author{A Baratin$^{1}$, B Dittrich$^2$, D Oriti$^2$ and J Tambornino$^3$}
\address{$^1$ Triangle de la Physique, CPHT \'Ecole Polytechnique, IPhT Saclay, LPT Orsay,\\
Laboratoire de Physique Th\'eorique, CNRS UMR 8627, Universit\'e Paris XI, F-91405 Orsay C\'edex, France}
\address{$^2$ Max Planck Institute for Gravitational Physics, Albert Einstein Institute,\\
	Am M\"uhlenberg 1, 14467 Golm, Germany}
\address{$^3$ Laboratoire de Physique, ENS Lyon, CNRS-UMR 5672,\\ 46 All\'ee d'Italie, Lyon 69007, France}

\begin{abstract}

The Hilbert space of loop quantum gravity is usually described in terms of cylindrical functionals of the gauge connection, the electric fluxes acting as non-commuting derivation operators. It has long been believed that this non-commutativity prevents a dual flux (or triad) representation of loop quantum gravity to exist. 
We show here, instead, that such a representation can be explicitly defined, by means of a non-commutative Fourier transform defined on the loop gravity state space.  In this dual  representation, flux operators act by $\star$-multiplication and holonomy operators act by  translation. 
We describe the gauge invariant dual states and discuss their geometrical meaning. 
Finally, we apply the construction to the simpler case of a $\U(1)$ gauge group and compare the resulting flux representation with the triad representation used in loop quantum cosmology.
\end{abstract}

\maketitle

\section{Introduction}

Loop quantum gravity (LQG) \cite{thiemannbook, rovellibook} is now a solid and promising candidate framework for a quantum theory of gravity in four spacetime dimensions. It is based on the canonical quantization of the phase space of general relativity in the Ashtekar formulation, using rigorous functional techniques as well as ideas and tools from lattice gauge theory. Diffeomorphism invariance of the classical theory is a crucial ingredient of the construction, both conceptually and mathematically, and background independence is the guiding principle inspiring it. The main achievement to date in this framework is the complete definition of the kinematical space of (gauge and diffeomorphism invariant) states of quantum geometry, based on the conjugate pair of variables given by holonomies $h_{e}[A]$ of the Ashtekar $\SU(2)$ connection $A$, and fluxes of the Ashtekar electric field $E$  (densitized triads) across 2-surfaces. These states are described in terms of so-called cylindrical functionals $\Psi[A]$ of the connection, which depend on $A$ via holonomies along graphs. Under suitable assumptions involving a requirement of diffeomorphism invariance, the representation of the algebra generated by holonomies and fluxes, hence the definition of the state space, is unique \cite{LOST}. 

A crucial, and somewhat surprising fact is that the flux variables, even at the classical level, do not (Poisson) commute \cite{ACZ, thiemann_qsd7}. This non-commutativity is generic and necessary, once holonomies of the Ashtekar connection are chosen as their conjugate variables. In the simplest case, for a given fixed graph, fluxes across surfaces dual to a single edge act as invariant vector fields on the group, and have the symplectic structure of the $\su(2)$ Lie algebra. Thus, the phase space associated to a graph is a product over the edges of cotangent bundles $T^\ast \SU(2) \simeq \SU(2) \times \su(2)$ on the gauge group.
For this case the Poisson structure for one edge $e$ (variables associated to different edges will commute) is simply given by
\be
\{h[A], h[A]  \} & = & 0 \nn \\
\{ E^i , h[A] \} & = &  \tau^i h[A] \nn \\
\{  E^i, E^j \} & = & -\epsilon^{ijk}E^k .
\ee
Here $E^i$ is the flux through an elementary (i.e. dual to a single edge $e$) surface $S_e$ with unit smearing function in a neighbourhood of the intersection point $e \cap S_e$\footnote{Note that we are working with rescaled flux variables. Thus, the Immirzi parameter $\gamma$ is implicitely hidden in the relation between Ashtekar's electric field and the triad through $E^a_j = \frac{1}{\gamma} \sqrt{\det q} e^a_j$. }. Recent works have shown that the structure of this phase space can also be understood from a simplicial geometric point of view \cite{biancajimmy, laurentsimone, valentin}. 

The fact that non-commutative structures are at the very root of the loop quantum gravity formalism is well-known for a long time\cite{ACZ}. However, to our knowledge, it has not been built upon to any extent in the LQG literature,  and the full implications of it, as well as the consequent links between the loop quantum gravity approach and non-commutative geometry ideas and tools, have remained unexplored. In fact,  it is often believed that non-commutativity of the fluxes implies that the framework has no flux (or triad) representation (for earlier attempts,  see for e.g \cite{Lewandowski}).  The goal of this paper is to show, instead, that this non-commutativity is naturally encoded in a definition of a non-commutative Fourier transform and $\star$-product, and that these can be used to build up a well-defined non-commutative flux representation for generic LQG states. 

The idea if defining a non-commutative flux representation for LQG originates from  developments in the spin foam context \cite{perez_spinfoams, daniele_thesis, baez_spinfoams},  and especially in the context of group field theory \cite{oriti_gft}. Much of the recent progress in spin foam models stemmed from the use of a coherent state basis \cite{eterasimone,eterasimone2, epr1,epr2,eprl,fk,asym} to express both quantum states and amplitudes. This basis has the advantage, as compared to the standard spin-network basis in LQG, of a clearer and more direct geometric interpretation  of the labels that characterize it, in terms of metric variables. This allowed a more consistent encoding of geometric constraints in the definition of the spin foam amplitudes, a nice characterization of the corresponding boundary states and of the semi-classical limit of the same amplitudes, relating them with simplicial gravity actions. The same aims also motivated recent work attempting to introduce metric variables in the group field theory framework \cite{generalisedGFT,generalisedGFT2}. 
This line of research has resulted in a new representation of group field theory in terms of non-commutative metric variables \cite{aristidedaniele}, which could in fact be directly interpreted as discrete (smeared) triads (in the $\SU(2)$ case). In this representation, where non-commutativity of metric variables is brought to the forefront and used in the very definition of the group field theory model, the Feynman amplitudes have the form of simplicial gravity path integrals in the same metric variables.
These results suggest to explore a similar metric representation for LQG states, since the group field can be interpreted as the (2nd quantized) wave function for a LQG spin network vertex. We exhibit such a representation here, and show that the whole construction of the LQG Hilbert space can be performed in this new representation as well.
  
We expect this new non-commutative flux representation to be useful in many respects. First of all it would help clarifying the quantum geometry of LQG states, including the relation with simplicial geometry \cite{biancajimmy, laurentsimone}. Thanks to this, it may facilitate the definition of the dynamics of the theory, both in the canonical (Hamiltonian or Master constraint) \cite{thiemannbook} and covariant (spin foam or GFT) setting \cite{aristidedaniele}, and the coupling of matter fields \cite{gtt2, danielehendryk,simone,aristide1,aristide2}. Further down the line, it offers a new handle for tackling the issue of the semi-classical limit of the theory. All these advantages of a metric representation are in fact shown already in the simpler context of Loop Quantum Cosmology, where such a representation has been already developed and used successfully \cite{bojowald_review, aps}. Obviously, the new representation brings loop quantum gravity closer to the language and framework of non-commutative geometry \cite{majid_book}, thus possibly fostering further progress.  

The paper is organized as follows. 
In section \ref{lqg_basis},  in order to make this paper self-contained, we review the standard construction of the kinematical Hilbert space of loop quantum gravity in the connection representation. The careful mathematical treatment of this review section will reveal useful for the rigorous construction of the new representation. 
In section \ref{FTsection}, we define the Fourier transform underlying the flux representation. The key technical ingredient is a generalization of the group Fourier transform \cite{PRIII, freidel_majid} to the whole LQG space of connections.  
In section \ref{fluxrep}, we describe further the new representation:  we give the action of the fundamental operators, we discuss properties of the gauge invariant dual states, clarifying their geometric meaning and the relation with the spin network basis. Finally, in section \ref{LQC}, we discuss the analogous construction in the simpler case of $\U(1)$ and comment on its relation with the triad representation used in Loop Quantum Cosmology. We conclude with a brief outlook on possible further developments.

\section{\label{lqg_basis}The Hilbert space of loop gravity}

Kinematical (gauge covariant) states in loop quantum gravity are functions on a space $\bar{\cA}$ of suitably generalized connections \cite{Velhinho}. 
A cornerstone of the framework is the fact that the state space $\cH_0$ can be defined by induction from a family of Hilbert spaces $\cH_{\gamma}\!=\!L^2(\cA_{\gamma}, \extd\mu_{\gamma})$,  labeled by graphs embedded in the spatial manifold $\sigma$. For a given graph $\gamma$ with $n$ edges,  $\cA_{\gamma}$ is a space of (distributional) connections on $\gamma$, naturally identified with the product $G^n$ of $n$ copies of the gauge group; $\extd\mu_{\gamma}$ is the product Haar measure on $G^n$.  
The construction stems from a characterization of $\bar{\cA}$ as a projective limit of the spaces $\cA_{\gamma}$. 

In this section we briefly recall this standard construction, as we will use it to  define the Fourier transform in section 3.  We will assume $G$ is any compact group, though having in mind the cases $G\!=\!\SU(2)$ or $\SO(3)$ relevant to gravity. Further details can be found in the original articles \cite{ashtekar_lewandowski_93, ashtekar_lewandowski_94_1, ashtekar_lewandowski_94_2} or in the textbook \cite{thiemannbook}.

\subsection{Generalized connections}

\label{A}

Given any smooth connection $A$ on $\Sigma$, one can assign a group element  $A_e$ to each path $e$ in $\Sigma$, by considering the holonomy  of $A$ along $e$. 
This assignment respects composition and inversion of paths:
\[
A_{e_1 \circ e_2} = A_{e_1} A_{e_2}, \qquad A_{e^{-1}} = A_{e}^{-1}.
\] 
In other words, the connection gives a morphism from the groupoid of paths to the gauge group $G$. 
The space $\bar{\cA}$ of  `generalized connections' is defined as the set ${\rm Hom}(\cP, G)$ of all such morphisms. 
It contains the smooth connections, but also distributional ones. $\bar{\cA}$  shows up as the quantum configuration space in loop quantum gravity. 

An independent and very useful characterization of $\bar{\cA}$ makes use of projective techniques \cite{Velhinho}, based on the set of embedded {\sl graphs}. 
A graph $\gamma\!=\!(e_1, \cdots, e_n)$ is a finite set of analytic paths with 1 or 2-endpoint boundary, such that every two distinct paths intersect only at one or two of their endpoints. The path components $e_i$ are called the edges of $\gamma$; the endpoints of an edge are called vertices. 
The set of all graphs has the structure of a partially ordered and directed set: we say $\gamma'$ is larger than $\gamma$, and we write $\gamma' \geq \gamma$, when  every edge of $\gamma$ can be obtained from a sequence of edges in $\gamma'$ by composition and/or orientation reversal; 
then for any two graphs $\gamma_1, \gamma_2$, there exists a graph $\gamma_3$ such that $\gamma_3 \geq \gamma_1, \gamma_2$.

For a given graph $\gamma$, let $\cA_{\gamma}\!:=\!{\rm Hom}(\bar{\gamma}, G)$ be the set of all morphisms from the subgroupoid $\bar{\gamma} \subset \cP$ generated by all the $n$ edges of $\gamma$, to the group $G$. 
$\cA_{\gamma}$ is naturally identified with $G^n$, both set-theoretically and topologically. For any two graphs such that $\gamma' \geq \gamma$, $\bar{\gamma}$ is a subgroupoid of $\bar{\gamma'}$: we thus have a natural  projection $p_{\gamma\gamma'} \maps \cA_{\gamma'} \to \cA_{\gamma}$, restricting  to $\gamma$ any morphism in $\cA_{\gamma'}$. These projections are surjective, and satisfy the rule:
\beq \label{projrule}
p_{\gamma\gamma'} \circ p_{\gamma'\gamma''} = p_{\gamma \gamma''}, \quad \forall \gamma'' \geq \gamma' \geq \gamma
\eeq
This defines a projective structure for the spaces $\cA_\gamma$. 
It can be shown \cite{ashtekat_lewandowski_94_2} that the space $\bar{\cA}$ coincides with the projective limit of the family ($\cA_{\gamma}, p_{\gamma\gamma'}$): namely, a generalized connection can be viewed as one of those elements 
$\{A_{\gamma}\}_{\gamma}$ of the direct product $\times_{\!\!\gamma} \cA_{\gamma}$ such that
\[
p_{\gamma\gamma'} A_{\gamma'} = A_{\gamma}, \quad \forall \gamma' \geq \gamma.
\]
Such a characterization allows to endow $\bar{\cA}$ with the topology of a compact Hausdorff space.

Let us close this section with a property of the projections $p_{\gamma\gamma'}$ that will be useful for us.  Given any two ordered graphs $\gamma' \geq \gamma$, the larger one $\gamma'$ may be obtained from the smaller one $\gamma$ by a sequence of three elementary moves: (i) {\sl adding} an edge (ii) {\sl subdividing} an edge by adding a new vertex (iii) {\sl inverting} an edge (see figure 1). 
\begin{figure}
\centering{
	\includegraphics[scale=1]{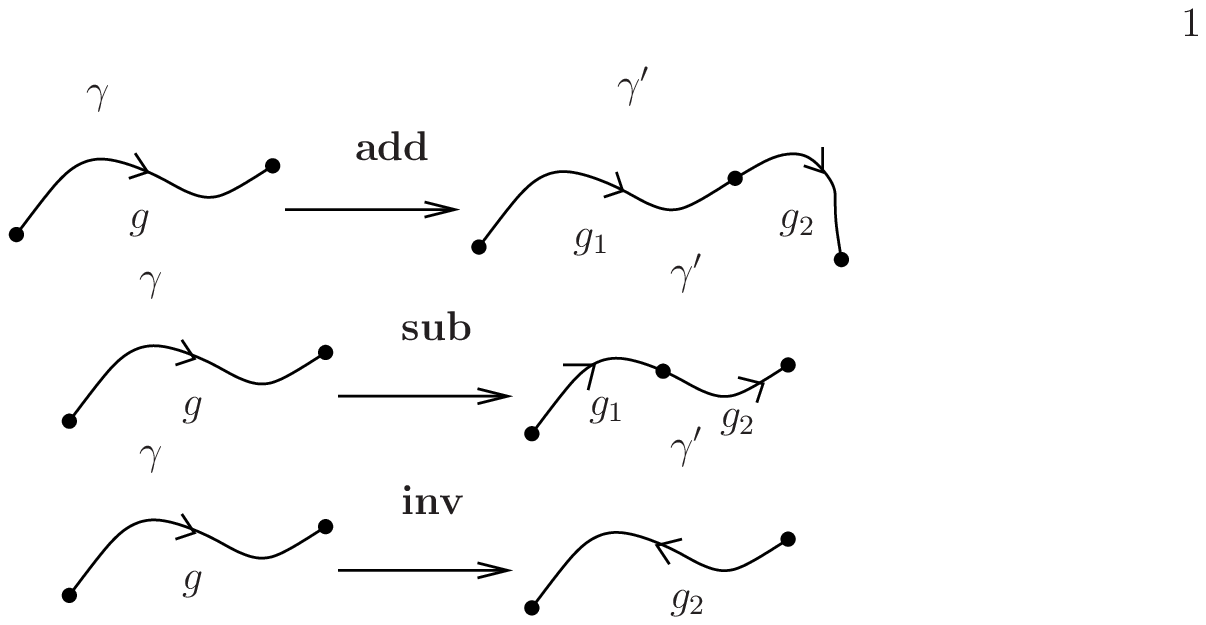}
\caption{Elementary moves relating ordered graphs}
}
\end{figure}
Together with the consistency rule (\ref{projrule}), this means that the projections $p_{\gamma\gamma'}$ can be decomposed into the following elementary projections onto the space $\cA_e$ of connections on a single edge $e$:
\be \nonumber
 p_{\rm add} & \maps & \cA_{e, e'} \rightarrow \cA_{e}; \,\,\quad (g, g')\mapsto g \\ \nonumber
p_{\rm sub} & \maps & \cA_{e_1, e_2} \rightarrow \cA_{e}; \quad (g_1, g_2) \mapsto g_1g_2\\
p_{\rm inv} & \maps & \cA_{e} \rightarrow \cA_{e}; \,\,\,\,\qquad  g \mapsto g^{-1}
\ee 
where we have used the identification $A_{\gamma} \!:=\! (g_1, \cdots g_n)$ of $\cA_{\gamma}$ with $G^n$.

\subsection{Inductive structure of $\cH_0$} 

Having understood the projective structure of the space of generalized connections:
 \[
 \bar{\cA} \simeq \{\{A_{\gamma}\}_{\gamma} \in \times_{\gamma} \cA_{\gamma}: \quad p_{\gamma \gamma'} A_{\gamma'} = A_{\gamma}  \quad \forall \gamma'\geq \gamma\},
 \]
we now illustrate how to define the LQG state space $\cH_0$ by an appropriate `glueing' of the much more tractable spaces $\cH_{\gamma}\!=\! L^2(\cA_{\gamma}, \extd \mu_{\gamma})$. The idea is to define functions on $\cA$ as equivalence classes of elements in $\cup_\gamma \cH_{\gamma}$ for a certain equivalence relation which reflects the projective structure of $\cA$. 

Let us introduce the family of injective maps $p^*_{\gamma'\gamma} \maps \cH_{\gamma} \to \cH_{\gamma'}$, $\gamma' \geq \gamma$, obtained by pull back of 
the projections $p_{\gamma\gamma'} \maps \cA_{\gamma'}\to \cA_{\gamma}$ defined in section \ref{A}. Thus $p^\ast_{\gamma'\gamma}$ acts on $f_{\gamma} \in \cH_{\gamma}$ as
\be
p^*_{\gamma'\gamma} \maps \cH_{\gamma} \to \cH_{\gamma'}, \quad (p^*_{\gamma' \gamma}f_{\gamma})[A_{\gamma'}] = f_\gamma[p_{\gamma \gamma'} A_{\gamma'}] 
\ee
These injective maps satisfy a rule analogous to (\ref{projrule}):
\be \label{injrule}
p^\ast_{\gamma''\gamma'} \circ p^*_{\gamma'\gamma} = p^\ast_{\gamma''\gamma}, \quad \forall \gamma'' \geq \gamma' \geq \gamma
\ee
Just as for the projections $p_{\gamma\gamma'}$,  the maps $p^\ast_{\gamma'\gamma}$ can be decomposed into three elementary injections 
$\add := p_{\rm add} ^*, \sub := p_{\rm sub}^*$ and $\inve :=
p^*_{\rm inv}$, which encode the transformation of the functions when adding, subdividing, and inverting an edge of a graph. 
These elementary injections act on the space $\cH_e$ associated to a single edge as:
\be \label{add_sub_1} 
&\add \maps &   \cH_{e} \rightarrow \cH_{e, e'}; \nn \\
& &  f(g) \mapsto (\add \acts f)(g, g') := f(g) \nn \\
& \sub \maps & \cH_{e} \rightarrow \cH_{e_1, e_2}; \nn \\
& &  f(g) \mapsto (\sub \acts f)(g_1, g_2) := f(g_1g_2) \nn \\
&\inve \maps & \cH_{e} \rightarrow \cH_{e}; \nn \\
& &  f(g) \mapsto (\inve \acts f)(g) := f(g^{-1}) \quad .
\ee
where we have  used once again the identification $A_{\gamma} \!:=\! (g_1, \cdots g_n)$ of $\cA_{\gamma}$ with $G^n$.  
Using these elementary maps, as well as the translation and inversion invariance and the normalization of the Haar measure, it can be checked that the $p^\ast_{\gamma\gamma'}$ are isometric embeddings $\cH_{\gamma} \hookrightarrow \cH_{\gamma'}$, namely injective maps preserving the inner product. This expresses the fact that $(\cH_{\gamma}, p^*_{\gamma'\gamma})_{\gamma'\geq \gamma}$ defines an inductive family of Hilbert spaces. 

We now define an equivalence relation on $\cup_{\gamma} \cH_{\gamma}$ by setting  
\[
f_{\gamma_1} \sim f_{\gamma_2} \quad \Longleftrightarrow \quad \exists \, \gamma_3 \geq \gamma_1, \gamma_2, \quad p^\ast_{\gamma_3\gamma_1} f_{\gamma_1} = p^\ast_{\gamma_3 \gamma_2} f_{\gamma_2}
\]
The quotient space can be endowed with an inner product which naturally extends the inner products $\langle\, , \, \rangle_{\gamma}$ of each $\cH_{\gamma}$. 
Let indeed $f_{\gamma_1}, f_{\gamma_2}$ be two functions in $\cup_{\gamma} \cH_{\gamma}$. The set of graphs is directed, so we may pick a graph $\gamma_3$ such that $\gamma_3 \geq \gamma_1, \gamma_2$.  It can then be easily shown using  the rule (\ref{injrule}) and the fact that the maps $p^*_{\gamma'\gamma}$ preserve the inner products, that the quantity
\[
\langle f_{\gamma_1} \,, \, f_{\gamma_2} \rangle := \langle p^\ast_{\gamma_3\gamma_1} f_{\gamma_1} \,,\, p^\ast_{\gamma_3\gamma_2} f_{\gamma_2} \rangle_{\gamma_3}
\]
does not depend on the chosen larger graph $\gamma_3$, and is well-defined on the equivalence classes $f_1\!:=\![f_{\gamma_1}]$ and $f_2\!:=\![f_{\gamma_2}]$. 
Hence it defines an inner product on the quotient space $\cup_{\gamma} \cH_{\gamma}/\!\!\sim$. The completion of this quotient space with respect to the inner product 
is called the inductive limit of the inductive family $(\cH_{\gamma}, p^*_{\gamma'\gamma})_{\gamma'\geq \gamma}$. 
It can be shown (see for example \cite{thiemannbook}) that the limit 
\be
\cH_{0} = \overline{ \cup_\gamma \cH_\gamma / \!\!\sim }
\ee 
coincides with the space $L^2(\bar{\cA}, \extd\mu_0)$ of square integrable functions on $\bar{\cA}$, with respect to a gauge and diffeomorphism invariant measure -- the so-called Ashtekar-Lewandowski measure \cite{ashtekar_lewandowski_94_2}. This is the kinematical (gauge covariant) state space of loop quantum gravity.

\subsection{Quantum theory on $\cH_0$}

Let us fix a graph $\gamma\!=\! (e_1, \cdots, e_n)$, and identify $\cH_{\gamma}$  with $L^2(G^n)$, where the $L^2$-measure is the product Haar measure. 
The fundamental operators arising from the quantization, on $\cH_{\gamma}$, of a classical phase space given 
by a cotangent bundle $T^\ast G^n$, act respectively by multiplication by a smooth function $\varphi_{\gamma}$ of $G^n$, and as generators of (right) actions of $G$ in (a dense subset of) $\cH_{\gamma}$:
 \be
(\vphihat_{\gamma} \acts f_{\gamma})(g_1, \dots, g_n) & := & \vphi_\gamma(g_1,
\dots, g_n) f_\gamma(g_1, \dots, g_n) \label{act_cyl_hol} \\ 
( \widehat{L}^i_e \acts f_{\gamma} )(g_1, \dots, g_n) & := &  \left. \frac{\extd}{\extd t} f_{\gamma}(g_1,
\dots, g_e e^{t\tau_i}, \dots, g_n)\right|_{ t=0} \label{act_cyl_flux}
\ee
where $\tau^i$ is a basis of $\su(2)$, say $i$ times the Pauli matrices, $\tau_i = i\sigma_i$. 
$\widehat{L}_e^i$ is the left-invariant vector field on the copy of $G$ associated to the edge $e$. 
This provides the quantum theory on the graph $\gamma$, with well-defined momenta operators, whose algebra has the structure of $\su(2)^n$. 

The action (\ref{act_cyl_hol}) can be easily extended to the quotient $\cup_{\gamma} \cH_{\gamma}/\!\!\sim$. For $\varphi_{\gamma_1}$  and $f_{\gamma_2}$ associated to different graphs, pick a graph $\gamma$ larger than both $\gamma_1$ and $\gamma_2$, and define $\vphihat_{\gamma_1} \acts f_{\gamma_2}$ as the equivalence class $[\vphihat_{\gamma} \acts f_{\gamma}]$ of (\ref{act_cyl_hol}). This action does not depend on the representatives chosen in the equivalence classes $\varphi:=[\varphi_{\gamma_1}]$ and $f:=[f_{\gamma_2}]$; it defines the action of the holonomy operator $\vphihat$ on generic states of $\cH_0$. 
The operator (\ref{act_cyl_flux}) should be interpreted as the flux $E_{S_e}^i \!:=\! E(S_e, \tau_i)$ of the electric field across an `elementary' surface\footnote{Actually there exist different proposals to which classical quantities the quantum flux operators should correspond: In \cite{thiemann_qsd7} it was shown that they can also be interpreted as quantum versions of a different set of classical functions involving the holonomies {\it and} the triads. The construction performed there is based on a family of graphs $\gamma$ {\it and} dual graphs $\gamma^*$ and the classical continuum phase space is understood as a certain generalized projective limit of graph--phase spaces of the form $T^*\SU(2)^n$. In section \ref{subsec:gauge_inv} we will see that this interpretation is also favored from the dual (Fourier transformed) point of view.} $S_e$ cut by the edge $e$. 
More generally, the LQG flux operator across a surface $S$ acts on $f=[f_{\gamma}]$ as a sum of left-invariant derivatives on $f_{\gamma'}$, where $\gamma' \geq \gamma$ cuts $S$ at its vertices, with only outgoing edges, the sum being over all the intersection points of $\gamma' \cap S$ and their adjacent edges:
\[
\widehat{E}^{i}_{S} \acts f_{\gamma}   = \sum_{v \in \gamma'\cap S}  \sum_{e \supset v} \, \epsilon(S, e) \, \widehat{L}^i_e \acts f_{\gamma},
\]
where $\epsilon(S, e) \!=\!\pm $ depends on the relative orientation of the edge and the surface. 

One can also define, on each $\cH_{\gamma}$, operators $\widehat{g_v}$ generating gauge transformations at each vertex of $v\in \gamma$.
These act  on a state $f_{\gamma}$ as
\beq \label{gauge}
(\widehat{g_v} \acts f_{\gamma})(g_1, \cdots, g_n) = f_{\gamma}(g^{\minus 1}_{s_1} g_1 g_{t_n}, \cdots, g^{\minus 1}_{s_n} g_{n} g_{t_{n}})
\eeq
where $s_e, t_e$ denote source and target vertices of the oriented graph $e$. Gauge invariance is thus imposed by acting with the gauge averaging operator 
\[
\cP_{\gamma} \!:=\! \bigotimes_v \int  \extd g_v\, \widehat{g}_v
\]
It can be checked that the action of such operators are well-defined on equivalence classes.  

Finally, the so called spin-network basis  of $\cH_{0}$ is a very convenient one for actual computations. Such a basis is obtained by harmonic analysis on the gauge group:
using the Peter--Weyl theorem, a state $f_\gamma \in \cH_\gamma$ can be decomposed into a product of Wigner functions $D^{j_e}_{m_e n_e}(g_e)$ for each edge, labeled by irreducible representations of $G$ ($j \in \frac{1}{2}\N$ for $G=\SU(2)$ or $j \in \N$ for $\SO(3)$), and magnetic numbers $-j_e \leq m_e \leq j_e$ and $-j_e \leq n_e \leq j_e$. 
These quantum numbers are usually interpreted as encoding geometric variables; in particular the spin $j$ labels the eigenvalues of area operators. 
In the next section, we define a Fourier transform on $\cH_0$ that will provide an alternative decomposition of the LQG states,  into functions of {\sl continuous} Lie algebra variables, naturally interpreted  as flux (triad)  variables.

\section{Fourier transform on the LQG state space}

\label{FTsection}

Here we define the non-commutative Fourier transform that will  give the dual flux representation. This transform generalizes the `group Fourier transform'   introduced in \cite{PRIII, freidel_majid, karim} to theories of connections.
We first recall the main features of the group  Fourier transform and use it to construct a family of Fourier transforms $\cF_\gamma$ defined on $\cH_{\gamma}$. 
We then show how this family extends to a transform $\cF$ defined on the whole space $\cH_{0}$. 
We emphasize that, to avoid unnecessary complications,  we will work from now on  with the gauge group $G\!=\!\SO(3)$. With more work,  the construction can be extended to the $\SU(2)$ case, using the $\SU(2)$ group Fourier transform spelled out in  \cite{karim}. 

\subsection{Group Fourier transform}

The $\SO(3)$ Fourier transform $\cF$ maps isometrically $L^2(\SO(3),\extd\mu_{H})$, equipped with the Haar measure $\extd \mu_{H}$, onto a space $L^2_{\star}(\R^3, \extd\mu)$ of functions on $\su(2)\! \sim \!\R^3$ equipped with a non-commutative $\star$-product, and the standard Lebesgue measure $\extd \mu$. 
 Just as for the standard Fourier transform on $\R^n$, the construction of $\cF$ stems from the definition of plane waves:
\[
\e_g \maps \su(2)\!\sim\! \R^3 \, \to\, \U(1) , \qquad \e_{g}(x) =e^{i \vec{p}_g\cdot \vec{x}}
\]
depending on a choice of coordinates $\vec{p}_g$ on the group manifold. For a given choice of such coordinates, $\cF$ is defined on $L^2(\SO(3))$ as 
\beq \label{FT}
\cF(f)(x) = \int \extd g f(g)\, \e_g(x)
\eeq
where $\extd g$ is the normalized Haar measure on the group\footnote{Since $\SO(3)$ is compact $L^2(\SO(3)) \subset L^1(\SO(3))$. Therefore the Fourier transform is well defined.}. 

Let us fix our conventions and notations. In the sequel we will identify functions of $\SO(3) \!\simeq\! \SU(2)/\Z_2$ with functions of $\SU(2)$ which are invariant under the transformation\footnote{Here $g$ is understood as an element of the fundamental representation of $G$. The transformation $g \rightarrow -g := h_\pi g$ can be understood as the action of an appropriate $h_\pi$, also also an element of the fundamental representation of $G$. } $g \to - g$. We denote by $\tau_i,\, i = 1, 2, 3$ the generators of $\su(2)$ algebra, chosen to be $i$ times the (hermitian) Pauli matrices. They are normalized as $(\tau_i)^2 \!=\! - \unit$ and satisfy $[\tau_i, \tau_j] \!=\! - 2 \epsilon_{ijk}\tau_k$. We choose  coordinates on $\SU(2)$ given by 
\[
\vec{p}_g = - \frac{1}{2}\Tr (|g| \vec{\tau}), \quad \quad |g| \! := \! \mbox{sign} (\Tr g)g
\]
where `$\Tr$' is the trace in the fundamental representation. The presence of the factor $\mbox{sign} (\Tr g)$ ensures that $\vec{p}_g\!=\!\vec{p}_{\minus g}$. 
Using these conventions, writing $x \!=\! \vec{x} \cdot \vec{\tau}$ and $g \!=\! e^{\theta \vec{n} \cdot \vec{\tau}}$ with $\theta\! \in \![0, \pi]$ and $\vec{n} \in \cS^2$, the plane waves take the form
\beq \label{planewaves}
\e_g(x)  =  e^{ - \frac{i}{2}\Tr (|g|x)} = e^{i\epsilon_{\theta}\sin \theta \vec{n} \cdot \vec{x}}
\eeq
with $\epsilon_{\theta} \!=\! \mbox{sign}(\cos\theta)$. Note that we may identify $\SO(3)$ to the upper hemisphere of $\SU(2)\!\sim\!\cS^3$, parametrized by 
$\theta \!\in\! [0, \pi/2]$ and $\vec{n}\!\in\!\cS^2$; on this hemisphere, we have $\epsilon_{\theta}=1$. 

The image of the Fourier transform (\ref{FT}) has a natural algebra structure inherited from the addition and the convolution product in $L^2(\SO(3))$. 
The product is defined on plane waves as
\beq \label{star}
\e_{g_1}\star \e_{g_2} = \e_{g_1g_2} \qquad \forall g_1, g_2 \in \SU(2)
\eeq
and extended by linearity to the image of $\cF$. Using the following identity 
\beq \label{deltagroup}
\int \extd^3 x \, \e_g(x)  = 4\pi[\delta_{\SU(2)}(g) +  \delta_{\SU(2)}(-g)] := 8\pi \, \delta_{\SO(3)}(g)
\eeq 
for the delta function on the group, with $\extd^3x$ being the standard Lebesgue measure on $\R^3$, one may prove the inverse formula
\[
f(g) =  \frac{1}{8\pi}\int \extd^3 x \, (\cF(f) \star \e_{g^\inv})(x),
\]
which shows that $\cF$ is invertible. Next, let us denote by $L^2_{\star}(\R^3)$ the image of $\cF$ endowed with the following Hermitian inner product:
\beq \label{innerproduct}
\langle u, v \rangle_{\star} := \frac{1}{8\pi} \int \extd^3 x (\overline{u} \star v) (x)
\eeq
Writing $u\!=\!\cF(f), v\!=\!\cF(h)$, the quantity $\langle u, v \rangle_{\star}$ can be written as:
\be
& & \langle \cF(f) ,\cF(h)\rangle_{\star} = \nn \\
& = &  \frac{1}{8\pi}\int \extd g_1 \extd g_2 \overline{f(g_1)} h(g_2) \int \extd^3 x (\overline{\e_{g_1}} \star \e_{g_2})(x)\nn \\
& = & \frac{1}{8\pi}\int \extd g_1 \extd g_2 \overline{f(g_1)} h(g_2) \int \extd^3 x\,  \e_{g^\inv_1 g_2}(x) \nn \\
& = &  \int \extd g  \overline{f(g)} h(g)  \nn
\ee
where on the second line we used that $\overline{\e_g(x)} = \e_{g^{\inv}}(x)$ as well as the identity (\ref{deltagroup}). This establishes in one stroke that 
the inner product (\ref{innerproduct}) is well defined, since $f$ and $h$ are square integrable, and that the Fourier transform defines a unitary equivalence $L^2(\SO(3))\!\simeq\!L^2_{\star}(\R^3)$. 

There are alternative ways to characterize the image $L^2_{\star}(\R^3)$ of the Fourier transform. To do so, we may recast the transform (\ref{FT}) into a standard $\R^3$ Fourier transform, in terms of the coordinates $\vec{p}_{g} \!=\! \sin\!\theta \vec{n}$, with $\theta\!\in\! [0, \pi/2]$.  Writing the Haar measure as $dg\!=\!\frac{1}{\pi} \sin^2\!\theta d \theta d^2\vec{n}$, where $d^2\vec{n} = \frac{1}{2} (\partial_i \vec{n} \times \partial_j \vec{n}, \vec{n}) dx^i \wedge dx^j, \;  (i,j \in \{ 1, 2 \})$ is the normalized measure on the unit sphere $\cS^2$, leads to the integral formula
\[
\cF(f)(x) = \frac{1}{\pi} \int_{|p| \leq 1} \frac{\extd^3\vec{p}}{\sqrt{1-p^2}} f(g(\vec{p})) e^{i \vec{p} \cdot \vec{x}} 
\]
We thus see that  the map $\cF$ hits functions of $\R^3$ that have {\sl bounded} Fourier modes $|\vec{p}_g| \leq 1$ for the standard $\R^3$ Fourier transform.  
We also may think of elements of $L^2_{\star}(\R^3)$ as equivalence classes of functions of $\R^3$, for the relation identifying two functions with the same $\R^3$-Fourier coefficients for (almost-every) low modes $|\vec{p}| \leq 1$. Loosely speaking, this means that the elements of $L^2_{\star}(\R^3)$ `probe' the space $\R^3$ with a finite resolution. 

It is worth noting that the image of the Fourier transform has a discrete basis, as shown by taking the Fourier transform of the Peter-Weyl formula: 
\beq \label{dualPW}
\widehat{f}(x) = \sum_{j, m, n} f^j_{mn}   \widehat{D}^j_{mn}(x)
\eeq
expressed in terms of the matrix elements of the dual Wigner matrices $\widehat{D}^j(x) \!=\! \int \extd g \e_g(x) D^j(g)$ in the $\SO(3)$ representation $j$. 

\subsection{Fourier transform on $\cH_{\gamma}$}

The extension of the above construction to functions of several copies of the group is straightforward, and gives the Fourier transform $\cF_{\gamma}$ on the space $\cH_{\gamma}\!\simeq\! L^2(\SO(3)^n)$  associated to any graph with $n$ edges.  
Given $\bold{g}\!:=\!(g_1, \cdots, g_n) \in \SO(3)^n$, we define the plane waves $E^{(n)}_{\bold{g}} \maps \su(2)^n \to \U(1)$ as a product of $\SO(3)$ plane waves:
\[
E^{(n)}_{\bold{g}}(\bold{x}) := \prod_{i=1}^n \, \e_{g_i}(x_i) 
\] 
The Fourier transform $\cF_{\gamma}$ is defined on  $\cH_{\gamma}$ by
\[
\cF_{\gamma}(f)(\bold{x}) =\int \prod_{i=1}^n \extd g_i  f(\bold{g}) \, E^{(n)}_{\bold{g}}(\bold{x}) 
\]
The $\star$-product acts on plane waves as
\[
(E^{(n)}_{\bold{g}} \star E^{(n)}_{\bold{g'}})(\bold{x}) := E^{(n)}_{\bold{g}\bold{g'}}(\bold{x}) =  \prod_{i=1}^n \, \e_{g_ig'_i}(x_i)  
\]
and is extended by linearity to the image of $\cF_{\gamma}$. This image, endowed with the inner product
\[
\langle u, v \rangle_{\star, \gamma} = \frac{1}{ (8\pi)^n}\int \prod_{i=1}^n \extd^3 x_i  \, (\overline{u} \star v)(\bold{x}),
\]
is a Hilbert space $L^2_\star(\R^{3})^{\!\otimes n}\!:=\!\cH_{\star,\gamma}$.  
The Fourier transform provides an unitary equivalence between the Hilbert spaces $\cH_{\gamma}$ and $\cH_{\star, \gamma}$.

\subsection{Cylindrical consistency and Fourier transform on $\cH_{0}$}

We have defined a family of unitary equivalences $\cF_{\gamma} \maps \cH_{\gamma} \to \cH_{\star, \gamma}$ labelled by graphs $\gamma$.
In this section we show the key technical result of this paper: this family extends to a map defined on the whole LQG state space 
\[
\cH_{0} = \overline{\cup_{\gamma} \cH_{\gamma} /\!\!\sim}\,\,.
\]
defined in section IIB. 

First, the family $\cF_{\gamma}$ gives a map $\cup_{\gamma} \cH_{\gamma} \to \cup_{\gamma} \cH_{\star, \gamma}$. 
In order to project it onto a well-defined map on the equivalence classes, we introduce the equivalence relation on $\cup_{\gamma} \cH_{\star, \gamma}$ which is `pushed forward' by $\cF_{\gamma}$: 
\[
\forall u_{\gamma_i} \in \cH_{\star, \gamma_i}, \quad u_{\gamma_1} \sim u_{\gamma_2} \quad \Longleftrightarrow \quad \cF^{- 1}_{\gamma_1}(u_{\gamma_1}) \sim \cF^{- 1}_{\gamma_2}(u_{\gamma_2})
\]
For simplicity, we use the same symbol $\sim$ for the equivalence relation in the source and target space. 
We thus have a map $\widetilde{\cF}$ making the following diagram commute:
\beq \label{diagram}
\xymatrix{
  \cup_{\gamma} \cH_{\gamma} \ar[r]^{\cF_{\gamma}} \ar[d]_{\pi} & **[r]\cup_{\gamma} \cH_{\star,\gamma} \ar[d]^{\pi_{\!\star}}
  \\
  \cup_{\gamma} \cH_{\gamma} /\!\!\sim  \ar[r]^{\widetilde{\cF}}& **[r]\cup_{\gamma} \cH_{\star,\gamma} /\!\!\sim
}
\eeq
where $\pi$ and $\pi_{\!\star}$ are the canonical projections. 
Next, the quotient space $\cup_{\gamma} \cH_{\star,\gamma} /\!\!\sim$ is endowed with a Hermitian inner product inherited from the inner products $\langle \, , \, \rangle_{\star, \gamma}$ on each $\cH_{\star, \gamma}$. This is also the inner product which is `pushed forward' by $\widetilde{\cF}$. 
The inner product of two elements $u, v$ of the quotient space with representatives $u_{\gamma_1}\!\in\!\cH_{\star,\gamma_1}$ and  $v_{\gamma_2}\!\in\!\cH_{\star,\gamma_2}$ is specified by choosing  a graph $\gamma_3 \!\geq\! \gamma_1, \gamma_2$ and two elements $u_{\gamma_3} \!\sim \! u_{\gamma_1}$ and $v_{\gamma_3} \!\sim\! v_{\gamma_1}$ in $\cH_{\star, \gamma_3}$, and by setting: 
\beq \label{big_innerproduct}
\langle u, v\rangle_{\star}  := \langle u_{\gamma_3},v_{\gamma_3} \rangle_{\star, \gamma_3}.
\eeq
In fact, we know by unitarity of $\cF_{\gamma_3}$ that the right-hand-side coincides with $\langle \cF^{-1}_{\gamma_3}(u_{\gamma_1}), \cF^{-1}_{\gamma_3}(v_{\gamma_2}) \rangle$, hence does not depend on the representatives $u_{\gamma_1}, v_{\gamma_2}$ nor on the graph $\gamma_3$.

It is worth giving a more concrete characterization of the space $\cup_{\gamma} \cH_{\star,\gamma} /\!\!\sim$, by making the equivalence relation and the inner product more explicit without using $\cF_{\gamma}$.  
As explained in section \ref{lqg_basis}, there are three generators of equivalence classes in $\cup_{\gamma} \cH_{\gamma}$, induced by the action on the set of graphs, consisting of adding, subdividing or changing the orientation of an edge.  These generators are encoded into the operators $\add$, $\sub$ and $\inve$ defined on $L^2(\SO(3))$. To characterize the equivalence classes in the target space, we thus need to compute the dual action of these operators on $L^2_{\star}(\R^3)$. We will need to introduce the following family of functions: 
\beq \label{delta}
\delta_x(y)  := \frac{1}{8\pi} \int \extd g \, \e_{g^\inv}(x)  \e_g(y) 
\eeq
These play the role of Dirac distributions in the non-commutative setting, in the sense that
\[
\int \extd^3 y \, (\delta_x \star f)(y) = \int \extd^3 y\, (f \star \delta_x)(y) = f(x) 
\]
However, let us emphasize that $\delta_x(y)$, seen as a function of $y\!\in\!\R^3$, is {\sl not} distributional;  this is a regular function\footnote{An explicit calculation using the expression (\ref{planewaves}) of the plane waves gives in fact 
$\delta_x(y) \!=\! \frac{1}{8\pi} \frac{J_1(|x-y|)}{|x-y|}$
where $J_1$ is the Bessel function of the first kind $J_n$ for $n\!=\!1$.} peaked on $y\!=\!x$, with a non-zero width, normalized as $\int \!\extd^3 y \, \delta_x(y) \!=\! 1$. We will denote by $\delta_0$ the function of this family obtained for the value $y\!=\!0$. 

Simple calculations show that the dual action of $\add$, $\sub$ and $\inve$ is given by:
\be
\label{add_sub_2}
\add \maps & & L^2_{\star}(\R^3) \rightarrow L^2_{\star}(\R^3)^{\!\otimes 2} \nn
\\
& &  (\add \acts u)(x_1,x_2) := 8 \pi u(x_1) \, \delta_0(x_2)\nn \\
\sub \maps & & L^2_{\star}(\R^3) \rightarrow L^2_{\star}(\R^3)^{\!\otimes 2} \nn \\
& &  (\sub \acts u)(x_1, x_2) := 8 \pi (\delta_{x_1} \star u)(x_2) \nn \\
\inve \maps & &  L^2_{\star}(\R^3) \rightarrow L^2_{\star}(\R^3) \nn \\
& &  (\inve \acts u)(x) := u( - x)\quad . \nn
\ee
Thus, when adding an edge, the function depends on the additional Lie algebra  variables $x_2$ via $\delta_0(x_2)$; taking the inner product of this function with any other function $v(x_1, x_2)$ of $L^2_{\star}(\R^3)^{\!\otimes 2}$ will project it onto its value $v(x_1, 0)$. When subdividing an edge into two parts, the two variables $x_1, x_2$ on the two sub-edges get identified (under inner product) via the function $\delta_{x_1}(x_2)$. Finally, when changing the orientation of the edge, the sign of the variable $x$ is flipped.

These rules describe recursively all the elements equivalent to $u$. By an obvious extension of these rules to functions on a graph with an arbitrary number of edges, they generate all the equivalence classes in  $\cup_{\gamma} \cH_{\star,\gamma}$.  It is instructive to check directly that the inner product given in (\ref{big_innerproduct}) is well-defined on equivalence classes. This amounts to showing that the linear maps $\add$, $\sub$ and $\inve$ acting on  $L_{\star}^2(\R^3)$ are unitary. For example, writing the inner product in $L_{\star}^2(\R^3)^{\!\otimes 2}$ as $\langle \, ,\, \rangle_{\star, 2}$, we easily check that, given $u, v \in L_{\star}^2(\R^3)$, we have
\be
& & \langle \add \acts u, \add \acts v \rangle_{\star, 2} = \nn \\
& = &  \int \extd^3 x_1 \extd^3 x_2 \, (\overline{u} \star v)(x_1) (\overline{\delta_0} \star  \delta_0)(x_2) \nn \\
& = & \langle u, v \rangle_{\star} \quad . \nn 
\ee
where the second equality follows from the fact that $\delta_0\!=\!\overline{\delta_0}$  is a $\star$-projector: $\delta_0 \star \delta_0 = \frac{1}{8\pi}\delta_0$, normalized to 1. Analogous calculations show the unitarity of $\sub$ and $\inve$. 

Coming back to the construction (\ref{diagram}), we now have a map $\widetilde{\cF} $ between two pre-Hilbert spaces, which, by construction, is invertible and unitary. 
Since $\cup_{\gamma} \cH_{\gamma}$ is dense in its completion $\overline{\cup_{\gamma} \cH_{\gamma}}$, there is a unique linear extension of $\widetilde{\cF}$ to a map
\[
\cF \maps \, \overline{\cup_{\gamma} \cH_{\gamma}/\!\sim} \,\, \longrightarrow \,\, \overline{\cup_{\gamma} \cH_{\star, \gamma}/\!\sim}
\]
between the completion of the two pre-Hilbert spaces. This defines our Fourier transform. $\cF$ is invertible and unitary, so that  it gives a unitary equivalence between 
the loop quantum gravity Hilbert space $\cH_{0}\!=\!\overline{\cup_{\gamma} \cH_{\gamma}/\!\sim}$ and the Hilbert space $\cH_{\star} \!=\! \overline{\cup_{\gamma} \cH_{\star, \gamma}/\!\sim}$.  

\section{Flux representation}
We now describe the representation obtained by applying the non--commutative Fourier transform onto the LQG state space. We derive the dual action of holonomy-- and flux-- operators, analyze the geometrical interpretation of this dual space and investigate its relation to the standard spin network basis.
\label{fluxrep}

\subsection{Dual action of holonomy and flux operators} 

\label{operators}

For a given fixed graph $\gamma$, consider an elementary  surface $S_e$ intersecting $\gamma$ at a single point of an edge $e$.  
The action of the flux operators $E^i_{S_e}$ on $\cH_{\gamma}$ coincides with the action (\ref{act_cyl_flux})  of left or right --invariant vector fields $\widehat{L}^i, \widehat{R}^i$ on $\SO(3)$, depending on the respective orientation  of $e$ and $S_e$ (see for example \cite{thiemannbook}). They act dually on $L^2_{\star}(\R^3)$ as $\widehat{L}^i \acts u\!:=\! \cF(\widehat{L}^i \acts f)$ and $R^i \acts u\!:=\! \cF(\widehat{R}^i \acts f)$, where $u\!=\!\cF(f)$. Now, since
\be
\cF(\widehat{R}^i\acts f)(x) & = & \int \extd g (\widehat{R}^i \acts f)(g) \e_g(x) \nn \\
& = & \int \extd g \left[\frac{\md}{\md t} f(e^{t\tau^i} g)\right]_{t=0} \e_g(x) \nn \\
& = &  \int \extd g f(g) \left[\frac{\md}{\md t} \e_{e^{\minus t\tau^i} g}(x) \right]_{t=0},\nn 
\ee
we only need to determine the action of the operators on the plane waves $e_g(x)$, for almost every $g$. By definition of the $\star$-product,  $\e_{e^{\minus t\tau^i} g} = \e_{e^{\minus t\tau^i}} \star  \e_g(x)$.  Thanks to the relation
\[
\left[\frac{\md}{\md t} \e_{e^{\minus t\tau^i}}(x) \right]_{t=0} = -\frac12\Tr(x \tau^i) =  -i x^i,
\]
we conclude that
$\widehat{R}^i \acts \e_g = -i \hat{x}^i \star  \e_g$, where $\hat{x}^i(x) \!=\! -\frac12\Tr(x \tau^i)$ is the coordinate function  on $\su(2)$. This shows that
\beq 
\cF(\widehat{R}^if)(x) = -i \hat{x}^i \star \cF(f) \quad .
\eeq
There is an analogous formula for the left--invariant vector field, which acts by $\star$-multiplication on the right.  Thus, the invariant vector fields on $\SO(3)$, and hence the elementary flux operator $E^i_{S_e}$ act dually by $\star$-multiplication. 

Next, we investigate the dual action of holonomy operators. 
We have seen that any
function $\vphi(g)$ 
defines a multiplication operators $\vphihat$ on $L^2(\SO(3))$.  
Let us consider the elementary operators $\widehat{\e}(a)$, labelled by Lie algebra variables $a \in \su(2)$,  generated by the plane waves $g \mapsto \e_g(a)$. 
Let $u\in L^2_{\star}(\R^3)$, and assume $u\!=\! \cF(f)$. The dual action of $\widehat{\e}(a)$ on $u$ is given by:
\[
(\widehat{\e}(a)\acts u)(x) := \cF(\widehat{\e}(a)\acts f)(x) = \int \extd g \, \e_g(a) f(g) \e_g(x) 
\]
Using the fact that $\e_g(a)\e_g(x) = \e_g(x+a)$, we obtain:
\[
(\widehat{\e}(a)\acts u)(x) = \cF(f)(x+a) = u(x+a)
\]
Hence elementary holonomy operators act by {\sl translation} on the states in the dual representations. More generally, any function $\vphi$ on the image $L^2_{\star}(\R^3)$ of the Fourier transform defines an operator $\vphihat$ acting on $f$ as 
\[
(\vphihat \acts f)(x) = \int \extd^3 a (\vphi \star_a f^x)(a)
\]
where $f^x(a)\!:=\!f(x+a)$.

As $\cF: \cH_0 \rightarrow \cH_\star$ is a unitary transformation, it preserves the spectra of all operators. In particular, geometrical quantities such as area or volume are quantized the same way as in the standard representation of loop quantum gravity. For instance, the area operator associated to an elementary surface $S_e$ is given by
\be
\hat{A}[S_e] := \gamma \sqrt{\delta_{ij} \hat{x}^i \star \hat{x}^j \star} ,
\ee
where the coordinate operators under the square root act by $\star$-multiplication and where the square root is defined via the spectral theorem. Note that, just as in the standard representation, we have the quantization ambiguity associated to Immirzi's parameter $\gamma$.

\subsection{Gauge invariant dual states}

\label{subsec:gauge_inv}

For a given graph $\gamma$, a gauge transformation at a vertex $v$ generated by a group element $g_v$ corresponds to the action of the operator $\widehat{g_v}$ on $\cH_{\gamma}$ given by Equ. \ref{gauge}. 
Consider a dual state $u_{\gamma}\!=\!\cF_{\gamma}(f_{\gamma})$, Fourier transform of a function $f_{\gamma}$.  The dual action of $\widehat{g}_v$ on $u_{\gamma}$ is defined as $\widehat{g}_v \acts u_{\gamma}\!:=\! \cF_{\gamma}(\widehat{g_v} \acts f_{\gamma})$ and read:
\[
(\widehat{g}_v \acts u_{\gamma})(x_{1}, \dots,  x_{n}) = u_{\gamma}(g^{\minus 1}_{t_1}x_{1} g_{s_1}, \dots,  g^{\minus 1}_{t_n} x_{n} g_{s_n})
\]
Gauge invariance is imposed by acting with the gauge averaging operator $\cP_{\gamma} \!:=\! \bigotimes_v \int \extd g_v\, \widehat{g}_v$.
The averaging over gauge transformation at a vertex $v$, assuming it has only outgoing edges, takes for the form: 
\[
(\int \extd g_v\, \widehat{g_v} \acts u_{\gamma}) (\bold{x}) = 
(\widehat{C}_v \star u)(x_{i}, \cdots x_{n})
\]
where $\widehat{C}_v$ is a `closure' constraint at the vertex $v$:
\[
\widehat{C}_v(x_{i}):= \int \extd g \prod_{e_i \supset v} \e_g(x_{i}) = 8 \pi \delta_0(\sum_{e_i \supset v} \, x_{i}) \quad .
\]
As emphasized in the previous section, the functions $\delta_0$ act as Dirac distribution for the $\star$-product; in particular $\delta_0 \star f \!=\! f \star \delta_0 \!=\! f(0) \delta_0$. Hence gauge invariance corresponds to  a {\sl strong} closure constraint for the $\su(2)$ variables $x_{i}$ of the edges incident at $v$.

More generally, the gauge invariant state $\cP_{\gamma} \acts u_{\gamma}$ is obtained by $\star$-multiplication of the function $u_{\gamma}$ with a product of closure constraints at each vertex $\widehat{C}_v\!=\!8 \pi\delta_0(\sum\limits_{e_i \supset v} \epsilon^i_vx_i)$, where $\epsilon_v^i \!=\!\pm 1$ depends on whether the edge $i$ is ingoing or outgoing at $v$. A nice way to write down a general expression for the gauge invariant states is the following. 
Consider the graph $\gamma' \geq \gamma$ obtained by (i) subdividing each edge $i \!\in\! \gamma$ into two parts $i_s, i_t$, where the sub-edge $i_s$ meet the `source' vertex $s_i$ and $i_t$ meet the `target' vertex $t_i$ of $i$; and (ii) by flipping the orientation of each $i_{t}$, so that the edges of the new graph $\gamma'$ are all outgoing of the original vertices of $\gamma$. This procedure defines a new element  $u_{\gamma'} \in \cH_{\gamma'}$ in the same equivalence class as $u_{\gamma}$, given by
\beq \label{refinedstate}
u_{\gamma'}(x_{1_s}, x_{1_t}, \cdots x_{n_s}, x_{n_t}) =(\prod_i \delta_{x_{i_s}} \star u_{\gamma})(\minus x_{i_t})
\eeq
Then the projector onto gauge invariant states acts on $u_{\gamma'}$ by left $\star$-multiplication by the product of closure constraints $\widehat{C}_v\! =\! 8 \pi \delta_0(\sum_{i_v\supset v} \, x_{i_v})$ at the vertices of $\gamma$:
\beq \label{projectorgauge}
\cP_{\gamma} \acts u = \left(\bigotimes_v \widehat{C}_v\right) \star u_{\gamma'}.
\eeq
The action of the projectors $\cP_{\gamma}$ is well defined on equivalence classes in $\cup_{\gamma}  \H_{\gamma}$; hence, by construction, it is also well-defined on the equivalence classes in  $\cup_{\gamma}  \H_{\star,\gamma}$. We may also check, directly from the definition (\ref{projectorgauge}), that the action of $\cP_{\gamma}$ commutes with the action of $\add, \sub$ and $\inve$. 

This only confirms the geometric interpretation of the Lie algebra variables $x_i$ as fluxes associated to elementary surfaces dual to the edges of the graph $\gamma$, and closing around vertices of the same graph to form elementary 3-cells\footnote{Note that the construction does not depend on the valence of the graph and thus does not need a simplicial setting for its geometric interpretation.}. To be more precise, it is useful to think of reference frames associated to the vertices of the graph $\gamma$. For a given state, the group Fourier variables $g_i$ associated to an oriented edge $i$ should be thought of as the parallel transport between the frames of the `source' and `target' vertices $s_i, t_i$. The flux variable $x_{i_s}$ (resp. $x_{i_t}$) in Equ. (\ref{refinedstate}) is then naturally interpreted as the flux across an elementary surface intersecting the edge $i$ at a single point, and then parallel-transported to the source vertex $s_i$ (resp. to  the target vertex $t_i$). These two flux variables, associated to the same edge, can then be identified with the relation $g_i x_{i_s} g_i^{-1}$. This relation is a consequence of the formula: 
\[
(\delta_{x} \star_y \e_{g})(y)= \e_{g} \star_x \delta_{g x g^\inv}(y)
\]
This geometrical interpretation is thus consistent with the action of plane waves and encoded into the star product.

\subsection{Relation with spin network basis}

It is interesting to investigate the relation between the Lie algebra variables $x$ and the labels of the standard basis of states. 
Starting from the geometric interpretation of $x$ as flux (or triad) variables, one would thus deduce from direct calculation the geometric interpretation of these labels. 
The relation with the usual spin-network basis is made explicit using the Fourier transform of the Peter-Weyl theorem, see Equ. \ref{dualPW}. 
This gives a basis for the dual states on a graph $\gamma$ given by a product over the edges of dual Wigner functions: 
\[
\widehat{D}_{m_en_e}^{j_e}(x) :=\int \extd g \, \e_g(x) D^{j_e}_{m_en_e}(g)
\]
These functions, whose dependence upon the norm $r\!=\!|x|$ of $x$ goes as  $J_{d_j}(r)/r$,  where $J_{d_j}$ is the Bessel function of the first kind associated to the integer $d_j\!:=\! 2j+1$ (see for e.g \cite{etera}), are peaked on the value $r\!=\!d_j$, thus relating  the spin $j$ to the norm of the flux.  
The quantum labels corresponding to the direction variables of the fluxes may then be identified using Perelomov group coherent states $| j, \vec{n}\rangle= g_{\vec{n}} | j , j\rangle$, where  $\vec{n}\in \mathcal{S}^2$ and $g_{\vec{n}}$ is an $\SU(2)$ element (say, the rotation with axis vector on the equator) mapping the north pole $(0,0,1)$ to  $\vec{n}$ by natural action on the  2-sphere $\mathcal{S}^2$. In such (overcomplete) coherent state basis,  the dual Wigner functions 
\[
\widehat{D}^{j}_{\vec{n}\vec{n}'}(x) := \langle j, \vec{n} | \widehat{D}^{j}(x) | j, \vec{n}'\rangle
\]
satisfy the property that 
\[
\widehat{D}^j_{\vec{n}\vec{n}'}(x) = \e_{g_{\vec{n}}g_{\vec{n}'}^\inv} \star \widehat{D}^j_{\vec{n}'\vec{n}'} = \widehat{D}^j_{\vec{n}\vec{n}} \star \e_{g_{\vec{n}}^\inv g_{\vec{n}'}} 
\]
where the diagonal matrix elements are given by
\beq  \label{coherentWigner}
\widehat{D}^j_{\vec{n}\vec{n}}(x)= \int \extd g \, \e_g(g_{\vec{n}}^{-1} x g_{\vec{n}}) \, D^j_{jj}(g)
\eeq
Now, the dependence of these function upon the directional part $\hat{x} \!=\! \vec{x}/|x|$ goes as $[\hat{x} \cdot \vec{n}]^{2j}$, and hence reaches its highest value for $\hat{x} \!=\! \pm \vec{n}$. 

These considerations suggest the identification $\vec{x} \!=\! j \vec{n}$ of flux variables and the labels of the coherent states basis, which should hold true in a suitable semi-classical limit. One can show that this is indeed the case, in the double limit  where fluxes and spins are large  $x \!\sim\! \frac{1}{\kappa}$, $j \!\sim\! \frac{1}{\kappa}$ with $\kappa \!\to\! 0$. This limit is obtained by introducing rescaled states $u_{\kappa}$ such that $u(x) \!=\! u_{\kappa}(\kappa x)$ and an effective $\star$-product $\star_{\kappa}$ making the rescaling unitary $\langle u, v\rangle_{\star} \!=\! \langle u_{\kappa}, v_{\kappa}\rangle_{\star_{\kappa}}$. Considering the modified plane waves:
\[
\e^{\kappa}_g(x) \!=\! e^{\frac{i}{\kappa}\epsilon_{\theta} \sin \theta \vec{n}}
\] 
where notations are the same as in (\ref{planewaves}), and the Fourier transform modified accordingly, this effective $\star$-product can be defined via its action on these plane waves as \[
\e^{\kappa}_{g_1} \star_{\kappa} \e^{\kappa}_{g_2}\!=\! e^{\kappa}_{g_1g_2}
\]  
By replacing $\e_g$ by $\e^{\kappa}_g$ in (\ref{coherentWigner}) and by rescaling the spins as $j \to j /\kappa$, one can then recast the integrand of the right hand side of (\ref{coherentWigner}) as an oscillatory phase, subject to saddle point analysis. The saddle point analysis is similar to the one performed in \cite{eterasimone}; we find that the existence of a saddle point requires precisely that $\vec{x} = j \vec{n}$. 

This confirms the interpretation of the spin $j$ as identifying eigenvalues of the (square of the) flux operators, thus  of their norm. In four dimensions, this gives areas to the elementary surfaces to which the flux variables are associated. We also conclude that, in the semi-classical limit, the coherent state parameters $\vec{n}$ behave like the direction components of the flux variables $\vec{x}$, and thus admit the same interpretation as triad components \footnote{This gives further support to the recent constructions in the spin foam setting \cite{epr1,fk, eterasimone,eterasimone2} based on group coherent states and on their interpretation as metric variables; in particular, it suggests that imposing geometric restrictions on them in the definition of the dynamical amplitudes will ensure that such amplitudes will have nice geometric properties in a semi-classical regime, as confirmed by the asymptotic analysis of \cite{asym}.}.

In general, therefore, we can expect that any function of the quantum numbers $j$,$\vec{n}$ will acquire, in a semi-classical approximation, a functional dependence on them matching that of the function $u(x)$ on the non-commutative triad variables $x$, in the same approximation\footnote{The asymptotic analysis of the new spin foam amplitudes \cite{asym}, showing how they take the form of a simplicial path integrals for gravity in the \lq\lq triad variables\rq\rq $j\vec{n}$ can then be interpreted as suggesting the existence (possibly beyond the semi-classical regime) of a simplicial path integral expression for the same amplitudes in the non-commutative variables $\vec{x}$. This interpretation is of course strongly supported by the results of \cite{aristidedaniele}.}.

\section{The $\U(1)$ case}

\label{LQC}

Here we shortly want to explain the Group Fourier transform for $\U(1)$
and comment on the relation to the triad representation used in Loop
Quantum Cosmology (see e.g. \cite{bojowald_review, aps}). The $\U(1)$
case is in several respects simpler than the $\SU(2)$ case but it can
serve to understand the principle mechanisms. As for $\SU(2)$ we start
by defining plane waves
\ba
\e_\phi(x) = e^{-i \phi x}
\ea
where $x\in \R$. The Fourier transform $\cF$ of a function $f(\phi)$ on
$\U(1)$ (with the convention $-\pi < \phi \leq \pi$) is then defined as
\ba\label{def1}
\cF(f) (x) &=& \int^\pi_{-\pi} \extd \phi   \, f(\phi)  \, \e_{\phi}(x) \nn \\
&=& \int^\pi_{-\pi} \extd \phi   \, f(\phi)  e^{-i \phi x}
\ea
Note the similarity with the usual Fourier transform which is obtained
by just restricting $x$ from $\R$ to $\Z$. The image  ${\rm Im}\cF$ is a
certain set of continuous functions on $\R$, but certainly not all
functions in $C(\R)$ are hit by $\cF$.\\
${\rm Im}\cF$ can be equipped with a $\star$--product, which is dual to
the convolution product on $\U(1)$. For plane waves, this product reads
\be
(\e_\phi\star \e_{\phi'})(x) := \e_{[\phi + \phi']}(x) \quad,
\ee
and extends to ${\rm Im}\cF$ by linearity. Here $[\phi + \phi']$ is the sum of the two angles modulus $2\pi$ such that  $-\pi<[\phi + \phi'] \leq   \pi$. In this way the star product is dual to group multiplication.
Next, we define an inner product on ${\rm Im}\cF$ via
\ba\label{innerp}
\langle u \,,\, v \rangle_{\star} := \int \extd  x \,\, (\overline{u}
\star v) (x) \quad \forall u,v \in {\rm Im \cF} \quad .
\ea
With this inner product one can check that $\cF$ is a unitary transformation between
$L^2(\U(1))$ and ${{\rm Im}\cF}$. \\
The peculiar class of functions which build up ${\rm Im}\cF$ also leads
to a different characterization of the $\star$--product: it turns out
that $\langle u, v \rangle_\star$ is entirely fixed by a discrete set of
values. This can be understood by comparing this Fourier transform with
the usual one which is obtained from (\ref{def1}) by restricting $x$ to
be integer, $x \in \Z$. In this case the inverse transformation is given
by
\ba\label{inversef}
f(\phi)= \frac{1}{2\pi}\sum_{x\in \Z}  \cF(f)(x) e^{i\phi x}  \q .
\ea
This formula indicates that for the function $u(x)$  in the image of
$\cF$, only the values $x\in \Z$ are relevant. Indeed we will see below
that the Lebesgue measure in $x$-space (together with the $\star
$--product) reduces to a counting measure with support in $\Z$ (and the
pointwise product) for functions $u \in {\rm Im}\cF$.\\
Using the formula for the inverse Fourier transform (\ref{inversef}),
the star product between two functions $u_1 \!=\! \cF(f_1)$ and $u_2\!=
\cF(f_2)$ can be evaluated to 
\be
& & u_1 \star u_2 \, (x) = \nn \\
& = & \int_{-\pi}^\pi  \int_{-\pi}^\pi  \extd \phi
\extd \phi' e^{-i\phi'x} \, f_1(\phi) f_2(\phi'-\phi) \nn\\
&=& 
 \sum_{x',x'' \in \Z}  u_1(x')
\, u_2(x'')   \frac{\sin(\pi(x'-x''))}{\pi(x'-x'')}
\frac{\sin(\pi(x''-x))}{\pi(x''-x)} \nn\\
&=&
\sum_{x'\in \Z}  u_1(x')
\, u_2(x')   \frac{\sin(\pi(x'-x))}{\pi(x'-x)}  
\ee
where for the last line we used that 
\ba
\frac{\sin(\pi(x'-x''))}{\pi(x'-x'')} =\delta_{x',x''}
\ea
for $x',x''\in \Z$. The integral over $x$ in
$\frac{\sin(\pi(x'-x))}{\pi(x'-x)}$ evaluates to one and therefore  the
inner product  (\ref{innerp}) is given by
\ba\label{ip2}
\langle u\,,\, v \rangle_{\star} = \int \extd  x \,\, (\overline{u}
\star v) (x) = \sum_{x\in \Z} \overline{u}(x)\, v(x) \q.
\ea
This agrees with the inner product for the usual Fourier transform.  As
mentioned the Lebesgue measure (to be understood together with the star multiplication) in the inner product (\ref{ip2})
can be rewritten as a counting measure (together with point multiplication) for functions $u \in {\rm Im}\cF$
which shows that we essentially have to deal with the Hilbert space of
square summable sequences, that is $\hat L_\star^2 (\R) \simeq \ell^2$.  With this counting measure there is a
large class of functions with zero norm inducing an equivalence relation
between functions that differ only by terms of zero norm, that is functions that are vanishing on all $x\in \Z$. In every
equivalence class one can define a standard representative by
\ba\label{sta}
u_s (x) =\sum_{x'\in \Z}  u (x')   \frac{\sin(\pi(x'-x))}{\pi(x'-x)} \	 .
\q
\ea 
These standard representatives also span  ${\rm Im}\cF$, that is, the condition $u \in {\rm Im}\cF$ picks a unique representative in the equivalence class.
Furthermore formula (\ref{sta})  defines the map that converts standard Fourier
transformed functions to group Fourier transformed functions and is in
precise analogy to the $\SU(2)$ case where we can use the `dual'
Peter--Weyl decomposition to show that functions in the image of $\cF$
can be sampled by discrete values.\\

On $L^2(\U(1))$ we have two elementary operators, the (left and right
invariant) derivative $L=-i\frac{\extd}{\extd \phi}$ and the holonomy
operator $T_n:=e^{-i\phi n}$, $n \in \Z$, that act as a multiplication operator. It
is straightforward to check, that these operators act dually as
\ba \label{ops1}
\hat L \acts u\,(x)&=& (x\star u )  \,(x) \nn\\
(\hat T_n \acts u)\,(x)&=& u(x+n)
\ea
In the same way as for $\SU(2)$ one can construct Hilbert spaces over
graphs and can also obtain cylindrical consistency of the group Fourier
transform map.

In Loop Quantum Cosmology (LQC) \cite{bojowald_review, aps}, a kind of
mini--superspace reduction of Loop Quantum Gravity, one uses also a
representation in which the (symmetry reduced) triad operator acts by
multiplication and the holonomies act by translations. The spectrum of
the multiplication operator is $\R$. Note that it is a discrete
spectrum in the sense that the associated eigenfunctions have finite
norm. This is possible as the Hilbert space used in LQC is
non-separable. Note that the representation (\ref{ops1}) used here is
different. The action of $\hat L$ is via $\star$-multiplication and  --
as in $L^2(\U(1))$ -- the spectrum is given by $\Z$.

The measure used in Loop Quantum Cosmology can be  defined through the inner product between two wave functions $u$ and $v$ in the following way.  Such a wave function $u$  can be understood as a map  from a countable set $\{x_i\}_{i\in I_{u}} \subset \R $ for some index set $I_{u}$ of countable cardinality to $\C$
\ba
u: x_i\rightarrow u(x_i)  \q .
\ea
The union of two countable sets $\{x_i\}_{i\in I_{u}}$  and  $\{x_i\}_{i\in I_{v}}$ defines another countable set which contains both previous sets.  In this way we obtain the structure of a partially ordered set similar to full Loop Quantum Gravity. Now one can extend each of the maps $u,v$ to the union of the two sets by defining $u(x):=0$ for all $x  \notin  \{x_i\}_{i\in I_{u}}$ and similarly for $v$. The inner product is given by
\ba\label{lqci}
\langle u \,,\, v  \rangle=  \sum_{x \in \{x_i\}_{i\in I_{u}} \cup \{x_i\}_{i\in I_{v}}}      \overline{u}(x) \, v(x)\q .
\ea
Hence wave functions  $u \in {\rm Im}\cF$ based on one copy of $U(1)$ can be (isometrically) embedded into the LQC Hilbert space, but the latter space is obviously much bigger.

\section{Outlook}

In this paper, we have used tools from non-commutative geometry, more precisely the non-commutative group Fourier transform of \cite{PRIII,freidel_majid,karim}, to define a new triad (flux) representation of Loop Quantum Gravity, which takes into account the fundamental non-commutativity of flux variables. We have shown first how this defines a unitary equivalent representation for states defined on given graphs (cylindrical functions), and then proven cylindrical consistency in this representation, thus defining the continuum limit and the full LQG Hilbert space. As one would expect, the new representation sees flux operators acting by $\star$-multiplication, while holonomies act as (exponentiated) translation operator. We have then discussed further properties of the new representation, including the triad counterpart of gauge invariance, clarifying further its geometric meaning and the relation with the spin network basis (including the case in which group coherent states are used). Finally, we have discussed the analogous construction in the simpler case of $\U(1)$ emphasizing similarities and differences with the triad representation commonly used in Loop Quantum Cosmology.

Let us conclude with a brief outlook on possible further developments.
As we mentioned in the text, our construction has been limited, for simplicity, to the case of $\SO(3)$ states. The extension of the group Fourier transform to $\SU(2)$ has been considered in \cite{freidel_majid, karim} and we expect the generalization of our construction of a LQG triad representation to be straightforward, and probably most easily performed using the plane waves augmented by polarization vectors (identifying the hemisphere in $\SU(2)$ in which the plane wave $\e_g(x)$ lives) defined in \cite{karim}. 

Perhaps more interesting is a fully covariant extension of the $\SU(2)$ structures we used to $\SO(4)$ or $\SL(2,\mathbb{C})$ ones, depending on the spacetime signature. In fact, we can think of our non-commutative triad vectors as identifying the self-dual or the rotation sector of the $\SO(4)$ or $\SL(2,\mathbb{C})$ algebra, and similarly for the group elements representing the conjugate connection. The $\SU(2)$ plane waves would then arise from $\SO(4)$ or $\SL(2,\mathbb{C})$ plane waves after imposition of suitable constraints corresponding to the constraints that reduce the phase space of BF theory to that of gravity, in a Plebanski formulation of 4d gravity as a constrained BF theory. It is at this level that the role of the Immirzi parameter (absent in our contruction) will be crucial. In identifying this covariant extension, one could take advantage of the detailed analysis of phase space variables and geometric constraints in \cite{biancajimmy}, in the simplicial context, and of the work already done on simplicity constraints in the non-commutative metric representation of GFTs in \cite{aristidedaniele}. This extension will most likely involve an embedding of the spatial $\SU(2)$ spin networks and cylindrical functions in spacetime obtained introducing unit vectors, interpreted as normals to the spatial hypersurface, located at the vertices of the graphs. The relevant structures would then be that of projected spin networks as studied in \cite{projectedSN, projectedSN2} (see also \cite{aristidedaniele}).  

As we mentioned in the text, our construction has identified the Hilbert space of {\it continuum} Loop Quantum Gravity in the new triad representation, by means of projective limits. It would be interesting, however, to obtain a better characterization of the resulting space in terms of some functional space of generalized flux fields, as we conjecture to be the case, in analogy to the usual construction of the $L^2$ space over generalized connections, endowed with the Ashtekar-Lewandowski measure. This will involve the definition of the relevant non-commutative $C^*$-algebra and the application of a generalization of the usual GNS construction (for some work in this direction, see \cite{lewandowski_okolow}). 

The new representation we have defined for LQG can be an important mathematical (and computational) tool for studying the semi-classical limit of the theory, using the expansion of the $\star$-product of functions in the Planck length (see \cite{PRIII}). In particular, this can be useful for a better understanding of quantum field theory for matter fields on a quantum spacetime, following \cite{gtt2}, and more generally for the definition of matter coupling in LQG. This is indeed already facilitated by the very presence of explicit triad (metric) variables in the quantum states of the theory, which is true in the new representation.

Finally, the new triad representation brings the geometric meaning of the LQG states to the forefront, and suggests a different avenue for the construction of coherent states, on top of giving of course a new representation for the known ones. Both these two facts can be relevant for tackling the issue of defining the quantum dynamics of the theory in the canonical framework, for analyzing the relation to the one defined by the new spin foam models \cite{epr1,epr2,eprl,fk}, and building up on the results of \cite{aristidedaniele} in the group field theory setting.


\begin{acknowledgements}
We thank Abhay Ashtekar for suggesting to analyze the relation between our construction and the Hilbert space used in loop quantum cosmology. Further we would like to thank Benjamin Bahr and Carlos Guedes for useful comments on a previous draft of this article. DO gratefully acknowledges financial support from the A. von Humboldt Foundation, through a Sofja Kovalevskaja Prize. JT is partially supported by an ANR ``Programme Blanc" grant LQG-09.
\end{acknowledgements}


\providecommand{\href}[2]{#2}\begingroup\raggedright\endgroup

\end{document}